\documentclass{cimento}

\usepackage{amssymb}
\usepackage{amsthm}
\usepackage{amsfonts}
\usepackage{graphicx}
\usepackage{amsmath}
\usepackage{amscd}
\usepackage{latexsym}

\input xy
\xyoption{all} \tolerance=500

\newtheorem{theorem}{Theorem}
\theoremstyle{definition}
\newtheorem{definition}[theorem]{Definition}

\newtheorem{example}[theorem]{Example}

\font\black=cmbx10 \font\sblack=cmbx7 \font\ssblack=cmbx5 \font\blackital=cmmib10  \skewchar\blackital='177
\font\sblackital=cmmib7 \skewchar\sblackital='177 \font\ssblackital=cmmib5 \skewchar\ssblackital='177
\font\sanss=cmss10 \font\ssanss=cmss8 
\font\sssanss=cmss8 scaled 600 \font\blackboard=msbm10 \font\sblackboard=msbm7 \font\ssblackboard=msbm5
\font\caligr=eusm10 \font\scaligr=eusm7 \font\sscaligr=eusm5  \font\fraktur=eufm10
\font\sfraktur=eufm7 \font\ssfraktur=eufm5

\font\bsymb=cmsy10 scaled\magstep2
\def\all#1{\setbox0=\hbox{\lower1.5pt\hbox{\bsymb
       \char"38}}\setbox1=\hbox{$_{#1}$} \box0\lower2pt\box1\;}
\def\exi#1{\setbox0=\hbox{\lower1.5pt\hbox{\bsymb \char"39}}
       \setbox1=\hbox{$_{#1}$} \box0\lower2pt\box1\;}

\def\tx#1{{\fam0\relax#1}}

\newfam\bifam
\textfont\bifam=\blackital \scriptfont\bifam=\sblackital \scriptscriptfont\bifam=\ssblackital

\newfam\blfam
\textfont\blfam=\black \scriptfont\blfam=\sblack \scriptscriptfont\blfam=\ssblack

\newfam\bbfam
\textfont\bbfam=\blackboard \scriptfont\bbfam=\sblackboard \scriptscriptfont\bbfam=\ssblackboard

\newfam\ssfam
\textfont\ssfam=\sanss \scriptfont\ssfam=\ssanss \scriptscriptfont\ssfam=\sssanss
\def\sss#1{{\fam\ssfam\relax#1}}

\newfam\clfam
\textfont\clfam=\caligr \scriptfont\clfam=\scaligr \scriptscriptfont\clfam=\sscaligr

\newfam\frfam
\textfont\frfam=\fraktur \scriptfont\frfam=\sfraktur \scriptscriptfont\frfam=\ssfraktur

\def\hpb#1{\setbox0=\hbox{${#1}$}
    \copy0 \kern-\wd0 \kern.2pt \box0}
\def\vpb#1{\setbox0=\hbox{${#1}$}
    \copy0 \kern-\wd0 \raise.08pt \box0}

\def\pmb#1{\setbox0\hbox{${#1}$} \copy0 \kern-\wd0 \kern.2pt \box0}
\def\pmbb#1{\setbox0\hbox{${#1}$} \copy0 \kern-\wd0
      \kern.2pt \copy0 \kern-\wd0 \kern.2pt \box0}
\def\pmbbb#1{\setbox0\hbox{${#1}$} \copy0 \kern-\wd0
      \kern.2pt \copy0 \kern-\wd0 \kern.2pt
    \copy0 \kern-\wd0 \kern.2pt \box0}
\def\pmxb#1{\setbox0\hbox{${#1}$} \copy0 \kern-\wd0
      \kern.2pt \copy0 \kern-\wd0 \kern.2pt
      \copy0 \kern-\wd0 \kern.2pt \copy0 \kern-\wd0 \kern.2pt \box0}
\def\pmxbb#1{\setbox0\hbox{${#1}$} \copy0 \kern-\wd0 \kern.2pt
      \copy0 \kern-\wd0 \kern.2pt
      \copy0 \kern-\wd0 \kern.2pt \copy0 \kern-\wd0 \kern.2pt
      \copy0 \kern-\wd0 \kern.2pt \box0}

\mathchardef\za="710B  
\mathchardef\zb="710C  
\mathchardef\zg="710D  
\mathchardef\zd="710E  
\mathchardef\zve="710F 
\mathchardef\zz="7110  
\mathchardef\zh="7111  
\mathchardef\zvy="7112 
\mathchardef\zi="7113  
\mathchardef\zk="7114  
\mathchardef\zl="7115  
\mathchardef\zm="7116  
\mathchardef\zn="7117  
\mathchardef\zx="7118  
\mathchardef\zp="7119  
\mathchardef\zr="711A  
\mathchardef\zs="711B  
\mathchardef\zt="711C  
\mathchardef\zu="711D  
\mathchardef\zvf="711E 
\mathchardef\zq="711F  
\mathchardef\zc="7120  
\mathchardef\zw="7121  
\mathchardef\ze="7122  
\mathchardef\zy="7123  
\mathchardef\zf="7124  
\mathchardef\zvr="7125 
\mathchardef\zvs="7126 
\mathchardef\zf="7127  
\mathchardef\zG="7000  
\mathchardef\zD="7001  
\mathchardef\zY="7002  
\mathchardef\zL="7003  
\mathchardef\zX="7004  
\mathchardef\zP="7005  
\mathchardef\zS="7006  
\mathchardef\zU="7007  
\mathchardef\zF="7008  
\mathchardef\zW="700A  
\mathchardef\zC="7009  

\newcommand{\be}{\begin{equation}}
\newcommand{\ee}{\end{equation}}

\newcommand{\bea}{\begin{eqnarray}}
\newcommand{\eea}{\end{eqnarray}}
\newcommand{\beas}{\begin{eqnarray*}}
\newcommand{\eeas}{\end{eqnarray*}}
\def\*{{\textstyle *}}
\newcommand{\R}{{\mathbb R}}

\def\ssT{\sss T}

\newcommand{\we}{\wedge}

\newcommand{\s}{{\textstyle *}}

\newcommand{\pa}{\partial}
\newcommand{\ti}{\times}

\def\cD{{\cal D}}

\def\cE{{\cal E}}

\def\cR{{\cal R}}

\def\cP{{\cal P}}

\def\sJ{{\sss J}}

\def\sT{{\sss T}}

\def\st{{\sss t}}

\def\sv{{\sss v}}

\def\sj{{\sss j}}

\def\xd{\tx{d}\,}
\def\xi{\tx{i}}

\def\cD{\cal D}

\newdir{ (}{{}*!/-5pt/@^{(}}

\def\eza{\za}
\def\ezb{\zb}

\def\ezw{\zw}

\def\ezvy{\zvy}

\def\xd{\operatorname{d}\!}

\def\s*{{\scriptstyle *}}

\newcommand{\bfr}{\begin{frame}}
\newcommand{\efr}{\end{frame} }

\newdir{|>}{%
!/4.5pt/@{|}*:(1,-.2)@^{>}*:(1,+.2)@_{>}}
\newdir{ (}{{}*!/-5pt/@^{(}}

\title{Tulczyjew triples in the constrained dynamics of strings \thanks{Research funded by the  Polish National Science Centre grant under the contract number DEC-2012/06/A/ST1/00256.}}
\author{J.~Grabowski\from{ins:x}\ETC,
K.~Grabowska\from{ins:y}
\atque
P.~Urba\'nski\from{ins:y}}
\instlist{\inst{ins:x} Institute of Mathematics,
                Polish Academy of Sciences - Warszawa, Poland
                \inst{ins:y} Faculty of Physics, University of Warsaw - Warszawa, Poland}

\PACSes{\PACSit{02.40}{Geometry, differential geometry, topology}
\PACSit{11.10.Ef}{Field theory, Lagrangian and Hamiltonian approach}}

\begin{document}

\maketitle

\begin{abstract}
We show that there exists a natural Tulczyjew triple in the dynamics of objects for which the standard (kinematic) configuration space $\textsf{T} M$ is replaced with $\wedge^n \textsf{T} M$. 
In this framework, which is completely covariant, we derive geometrically phase equations, as well as Euler-Lagrange equations, including nonholonomic constraints into the picture. Dynamics of strings and a constrained Plateau problem in statics are particular cases of this framework.
\end{abstract}
\section{Introduction}
This work is a continuation of a research undertaken jointly with W.~M.~Tulczyjew on the Legendre transformation in the dynamics of strings.
We show that there exists a natural Tulczyjew triple in the dynamics of objects for which the standard (kinematic) configuration space $\textsf{T} M$ is replaced with $\wedge^n \textsf{T} M$. To do that, we make use of graded bundles of degree $n$, i.e. objects generalizing vector bundles (for which $n=1$). For instance, the role of $\textsf{T}\textsf{T}^\*M$ is played in our approach by the manifold $\we^n\textsf{T} M\we^n\textsf{T}^\*M$, which is canonically a graded bundle of degree $n$ over $\we^n\textsf{T} M$. To obtain the dynamics, we use the canonical multisymplectic $n+1$-form $\ezw^n_M$ on $\wedge^n\sT^\* M$, which gives, by the contraction, canonical morphism $$\ezb^n_M \colon \wedge^n \sT\wedge^n\sT^\* M \rightarrow \sT^\*\wedge^m\sT^\* M \,.
$$
The dynamics of strings and the Plateau problem in statics are particular cases of this framework.
We refer to our work \cite{GGU3} if details of this concept and references are concerned. Here, we add nonholonomic constraints into the picture together with a geometric description of the corresponding d'Alembert principle and constrained Euler-Lagrange equation. Sine we are working with Tulczyjew triples, there is no need to introduce Poincar\'e-Cartan forms and the presentation of nonholonomic constraints seems to be simpler than that known from the literature (c.f. \cite{BLMS,VCLM}, although the idea of d'Alembert principle is preserved.
The presented approach using multivectors can be viewed also as a simplified version of a variational calculus for non-parameterized submanifolds, whose full version is technically much more complicated (see e.g. \cite{Lu}). Some Tulczyjew triples for general classical field theories appeared recently in \cite{G1,G2,GG2}.

We want to stress that our framework is completely covariant and, what is important, not reduced to derivation of just the Euler-Lagrange equations.
We present the full picture, determining the phase space and the phase equations as the meeting point of the Lagrange and the Hamilton formalisms, subject to the corresponding \emph{Legendre transformation}. The equations are obtained purely geometrically by means of the morphisms in the triple from the Lagrangian submanifolds generated by Lagrangians or Hamiltonians. In particular, on the Hamiltonian side we do not use, at least explicitly, any Poisson brackets.
For the basics on Tulczyjew triples and Legendre transformations we reccomend \cite{Tu1,Tu6,Tu3}.

Note finally that classical field theory is usually associated with the concept of a multisymplectic structure.  The multisymplectic approach appeared first in the papers of the `Polish school' \cite{Ga,KS,KTu,Tu5}. Then, it was developed by Gotay, Isennberg, Marsden, and others in \cite{GIMa,GIMb}. The original idea of the multisymplectic structure has been thoroughly investigated and developed by many contemporary authors, see e.g. \cite{CIL,CCI1,EM,FP1,FP2,RR}. The Tulczyjew triple in the context of multisymplectic field theories appeared recently in \cite{CGM} and \cite{LMS} (see also \cite{V}). A similar picture, however with differences on the Hamiltonian side, one can find in \cite{GM} (see also \cite{GMS,Kr}) and many others; it is not possible to list all achievements in this area.

Note, however, that the multisymplectic structure which appears in this paper is a canonical structure on $\we^n\sT^*M$ and the question about the proper abstract generalization of this structure is not discussed here.
On the other hand, it would be interesting to develop a similar theory for more general objects in the spirit in which mechanics on (Lie) algebroids generalizes the classical one for $\sT M$
(see e.g. \cite{GG,GGU2,Mar,We}).

\section{The standard Tulczyjew triple}

The canonical symplectic form $\zw_M$ on $\sT^\* M$ induces an isomorphism
$$\zb_M:\sT\sT^\* M\to\sT^\*\sT^\* M\,.$$
Composing it with $\cR_{\sT M}$, where
$$\cR_E:\sT^\*E^\*\to \sT^\*E$$
is the well-known canonical isomorphism o double vector bundles (see e.g. \cite{KU,Mc,Ur}),
 we get the map
$$\za_M:\sT\sT^\* M\to\sT^\*\sT M\,.$$
Using the standard coordinates $(x^\zm,\dot{x}^\zn)$ and $(x^\zm,p_\zn)$ on $\sT M$ and $\sT^*M$, respectively, and the adapted coordinates on $\sT^*\sT M$ and $\sT\sT^*M$, we can write
\be\label{alpha} \za(x,p,\dot x,\dot p)=(x,\dot x,\dot p,p)\,.
\ee
This gives rise to the commutative diagram of \emph{double vector bundle (iso)morphisms} (Tulczyjew triple)
\be\label{tt}
    \xymatrix@R-4mm @C-10mm
        { & \sT^\*\sT^\*M \ar[ldd]_*{} \ar[rd]^*{} & & & \sT \sT^\* M \ar[rrr]^*{{\eza}_M}
        \ar[lll]_*{\ezb_M} \ar[ldd]^*{} \ar[rd]^*{}& & & \sT^\*\sT M \ar[ldd]^*{} \ar[rd]^*{} & \cr
        & & \sT M \ar[ldd]^*{} & & & \sT M  \ar[ldd]^*{} \ar[lll]^*{} \ar[rrr]^*{} & & & \sT M \ar[ldd]^*{}  \cr
         \sT^\* M  \ar[rd]^*{}  & & & \sT^\* M \ar[rrr]^*{} \ar[lll]^*{} \ar[rd]^*{}& & & \sT^\* M
        \ar[rd]^*{} & & \cr
        & M  & & &  M \ar[rrr]^*{} \ar[lll]^*{} & & & M & }.
\ee

Note that the mapping $\za_M$ can be obtained directly as the dual to the `canonical flip' $\zk_M:\sT\sT M\to \sT\sT M$,
 which is an isomorphism of two vector bundle structures on $\sT\sT M$:
\be\label{kappa}
    {\xymatrix@R-3mm @C-2mm{  & \sT \sT M \ar[ldd]_*{\sT\zt_M} \ar[rd]^*{\zt_{\ssT M}}
    \ar[rrr]^*{\zk_M} & & & \sT \sT M \ar[ldd]_(.3)*{\zt_{\ssT M }}  \ar[rd]^*{\sT\zt_M} & \cr
    & & \sT M \ar[ldd]_(.3)*{\zt_M}  \ar[rrr]^(.7)*{\text{id}} & & & \sT M \ar[ldd]_*{\zt_M} \cr
    \sT M  \ar[rd]^*{\zt_M} \ar[rrr]^(.7)*{\text{id} } & & &
    \sT M \ar[rd]^*{\zt_M}  & &  \cr
    & M \ar[rrr]^*{\text{id}} & & & M &}} .\ \ \
\ee
Indeed, the duals of these two vector bundle structures on $\sT\sT M$ are $\sT^\*\sT M$ and $\sT\sT^\* M$, and $\za_M$ can be understood as the dual map of $\zk_M$.

\medskip\noindent
The map $\zk_M$, as well as $\za_M$ and $\zb_M$, encodes the Lie algebroid structure of $\sT M$ and note that no brackets are needed (cf. \cite{GU3,GU1}).

\section{The Tulczyjew triple - Lagrangian and Hamiltonian formalisms}
Let now, for a mechanical system, $M$ be the manifold of positions, so that
the tangent bundle $\sT M$ represents (kinematic) configurations and $\sT^\ast M$ is the phase space, and let
$L:\sT M\rightarrow \R$ be a Lagrangian function.
Putting all this into the Tulczyjew triple, we get the diagram
{$$\hskip-1cm\xymatrix@C-20pt@R-10pt{
{\mathcal{D}}\ar@{ (->}[r]& \sT\sT^\ast M \ar[rrr]^{\alpha_M} \ar[dr]\ar[ddl]
 & & & \sT^\ast\sT M\ar[dr]_{\pi_{\sT M}}\ar[ddl] & \\
 & & \sT M\ar@{.}[rrr]\ar@{.}[ddl]
 & & & \sT M \ar@{.}[ddl]\ar@/_1pc/[ul]_{\xd L}\ar[dll]_{\cP L}\\
 \sT^\ast M\ar@{.}[rrr]\ar@{.}[dr]
 & & & \sT^\ast M\ar@{.}[dr] & &  \\
 & M\ar@{.}[rrr]& & & M &
}\qquad$$}
Here,
{$$\cP L:\sT M\rightarrow \sT^\ast M, \;\; \quad \cP L\left(x,\dot x\right)=
(x,\frac{\partial L}{\partial \dot x})$$}
is the \emph{Legendre map}, and the submanifold
\be\label{Le}\mathcal{D}=\alpha_M^{-1}(\xd L(\sT M)))
\ee
of $\sT\sT^*M$ represents the (implicit in general) \emph{Lagrange (phase) equation}.
Recall that, by definition, a curve $\zb:\R\to N$ is a solution of an implicit differential equation (differential relation) $\mathcal{D}\subset\sT N$ if its tangent prolongation $\st\zb:\R\to\sT N$ takes values in $\cD$. In local coordinates,
{$$\mathcal{D}=\left\{(x,p,\dot x,\dot p):\;\; p=\frac{\partial L}{\partial \dot x},\quad \dot p=\frac{\partial L}{\partial x}\right\}\,,$$
so that the phase equations in an implicit form read
$$p=\frac{\partial L}{\partial \dot x},\quad \dot p=\frac{\partial L}{\partial x}\,.$$


The Hamiltonian formalism looks analogously. If {$H:\sT^\ast M\rightarrow \R$} is a Hamiltonian function, from the Hamiltonian side of the triple
{$$\hskip-1.2cm
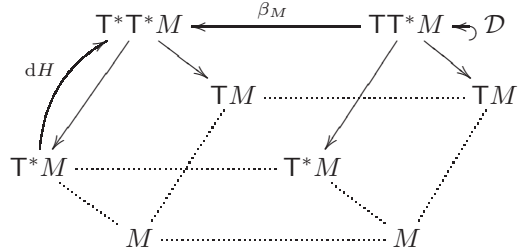
\xymatrix@C-20pt@R-10pt{
 & \sT^\ast\sT^\ast M  \ar[dr] \ar[ddl]
 & & & \sT\sT^\ast M\ar[dr]\ar[ddl] \ar[lll]_{\beta_M}&
 { \mathcal{D}}\ar@{ (->}[l] \\
 & & \sT M\ar@{.}[rrr]\ar@{.}[ddl]
 & & & \sT M \ar@{.}[ddl]\\
 \sT^\ast M\ar@{.}[rrr]\ar@{.}[dr] \ar@/^1pc/[uur]^{\xd H}
 & & & \sT^\ast M\ar@{.}[dr] & &  \\
 & M\ar@{.}[rrr]& & & M &
}\qquad$$}
we derive the phase dynamics in the form
{$$\mathcal{D}=\beta_M^{-1}(\xd H(\sT^\ast M))\,.$$}
 This dynamics is automatically explicit, i.e. generated by the Hamiltonian vector field, so it corresponds to a phase dynamics induced by a Lagrangian function only in regular cases. In general, one has to use more sophisticated tools like \emph{Morse families} etc., see \cite{TU}.
In local coordinates,
{$$\mathcal{D}=\left\{(x,p,\dot x,\dot p):\;\; \dot p=-\frac{\partial H}{\partial x},\quad \dot x=\frac{\partial H}{\partial p}\right\}\,,$$}
so we obtain the standard Hamilton equations.

\section{Nonholonomic constraints and Euler-Lagrange equations}
Let now, $\zg:\R\to M$ be a curve in $M$ (of course, $\R$ can be replaced by an open interval), and $\st\zg:\R\to\sT M$ be its tangent prolongation. It is easy to see that both curves, $\xd L\circ\st\zg$ and $\za_M\circ\st(\cP L\circ\st\zg)$ are curves in $\sT^*\sT M$ covering $\st\zg$. Therefore, their difference makes sense and, as easily seen, takes values in the annihilator $V^0\sT M$ of the vertical subbundle $V\sT M\subset\sT\sT M$. Since $V^0\sT M\simeq\sT M\ti_M\sT^*M$, we obtain a map
$\zd L_\zg:\R\to\sT^*M$. This map is usually interpreted as the external force along the trajectory. Its value at $t\in\R$ depends only on the second jet $\st^2\zg(t)$ of $\gamma$, so defines the variation of the Lagrangian understood as a map
\be\label{work}\zd L:\sT^2M\to\sT^*M\,,
\ee
where $\sT^2M$, the second tangent bundle, is the bundle of all second jets of curves $\R\to M$ at $0\in\R$.
The equation
\be\label{EL}\zd L_\zg=\zd L\circ\st^2\zg=0
\ee
is known as the \emph{Euler-Lagrange equation} and tells that the curve $\xd L\circ\st\zg$ corresponds \emph{via} $\za_M$ to an \emph{admissible curve} in $\sT\sT^*M$, i.e. the tangent prolongation of a curve in $\sT^*M$. Here, of course, $\st^2\zg$ is the second tangent prolongation of $\zg$ to $\sT^2M$.

If now, $A\subset\sT M$ is an affine subbundle of $\sT M$, $\sv(A)$ is the linear part of $A$, and $\sv(A)^0\subset\sT^*M$ its annihilator, then
we can replace (\ref{EL}) with
\be\label{dAprinciple}\zd L_\zg\in \sv(A)^0\,,
\ee
which is the \emph{d'Alembert principle}. It tells that the forces $\zd L_\zg$ belong to $\sv(A)^0$, so make no work along the trajectory.
The constrained \emph{nonholonomic Euler-Lagrange equations} associated with the affine nonholonomic constraint represented by $A$ take the form
\bea\label{cEL} &&\st\zg\in A\,;\\
\label{cEL1} && \zd L_\zg\in \sv(A)^0\,.
\eea
In a more traditional form, they can be expressed in local coordinates as
\bea\label{nhEL} &&\left(\dot x^\zs(t)-a^\zs)\right)\zh^i_\zs(x(t))=0\,;\\
&&\frac{\partial L}{\partial x^\zs}-\frac{\xd}{\xd
t}\left(\frac{\partial L}{\partial \dot{x}^\zs}\right)=\zl_i\zh^i_\zs\,,
\eea
where $\zh^i=\zh^i_\sigma(x)\xd x^\sigma$, $i=1,\dots,\dim(\sv(A)^0)$, are one-forms generating
$\sv(A)^0$, $a(x)=(x^\zm,a^\zs(x))$ is an arbitrary section of the affine bundle $A\to M$, and $\zl_i(x)$ are arbitrary coefficients, traditionally interpreted as "Lagrange multipliers".
The section $a$ can be chosen $0$ if the constraints are actually linear.


\section{The Tulczyjew triple with the kinematic configuration space $\we^n\top M$}

We want to build a similar framework replacing points with higher dimensional objects, being motivated by the study of dynamics of one-dimensional non-parameterized  objects  (strings).

The \emph{motion} of a system  will be given by an $n$-dimensional submanifold in the manifold $M$ (``space-time'').  An infinitesimal piece of the motion is the first jet of the submanifold.
However, this model leads to essential complications even in one-dimensional case (relativistic particle).  For instance, the infinitesimal action (Lagrangian) is not a function on first jets, but a section of certain line bundle over the first-jet manifold, a `dual' of the bundle of ``first jets with volumes''.
Therefore we will take the compromise and use for the space of infinitesimal pieces of motions the space of simple $n$-vectors, which represent first jets of $n$-dimensional submanifolds together with an infinitesimal volume.  It is technically convenient to extend this space to all $n$-vectors, i.e. to  the vector bundle $\wedge^n\sT M$ of $n$-vectors on $M$.
In this way we get the following principles:
\begin{itemize}
\item A \emph{Lagrangian} $L$ is a function on infinitesimal motions, $L:\wedge^n\sT M\to\R$.
 If $L$  is positive homogeneous, the action functional does not depend on the parametrization of the submanifold and the corresponding Hamiltonian (if it exists) is a function on the dual vector bundle  $\wedge^n\sT^\* M$ (the phase space).
\item The \emph{dynamics} should be an equation (in general, implicit) for $n$-dimensional submanifolds in the phase space, i.e.
    $${\cD}\subset\wedge^n \sT \wedge^n\sT^\* M\,.$$
\item A submanifold $S$ in the phase space $\wedge^n\sT^\* M$ is a \emph{solution} of $\cD$ if and only if its tangent space $\sT_\za S$ at $\za\in\wedge^n\sT^\* M$ is represented by a $n$-vector from ${\cD}_\za$. If we use a parametrization, then the tangent $n$-vectors associated with this parametrization must belong to $\cD$.
\end{itemize}

For simplicity, in what follows we will consider the `string case' $n=2$, but the constructions remain valid for arbitrary $n$. We will use canonical coordinates $(x^\zr,\dot x^{\zm\zn})$ and $(x^\zr,p_{\zm\zn})$ on $\we^2\sT M$ and $\we^2\sT^*M$ (with the convention
    $\dot x^{\zm\zn}=-\dot x^{\zn\zm}$, $p_{\zm\zn}=-p_{\zn\zm})$, respectively, representing the decomposition of bivectors:
    $$\dot x^{\zm\zn}\pa_{x^\zm}\we\pa_{x^\zn}\in\we^2\sT M\,,\quad p_{\zm\zn}\xd x^\zm\we\xd x^\zn\in\we^2\sT^*M\,.$$
Since $\wedge^n \sT \wedge^n\sT^\*M$ is NOT a double vector bundle for $n>1$, we start with introducing objects naturally generalizing vector bundles (see \cite{GR,GR2}).

\begin{definition}
A \emph{graded bundle} of degree $r$ is a fibration  $\cE\to M$ such that the typical fiber is $\R^k$, with coordinates $(y^1,\dots,y^k)$ which have associated \emph{weights} (or \emph{degrees}), $w_1,\dots,w_k=1,\dots,r$, \  respected by the fiber-bundle change of coordinates.
\end{definition}

We extend weights in fibres, associating weights $0$ with basic functions, thus having local coordinates $(x^\zm,y^a)$ consisting of homogeneous functions.
Such a structure can be conveniently encoded by the \emph{weight vector field}
$$X_\cE=\sum_aw_ay^a\pa_{y^a}\,,$$
whose flow extends to an action $\R\ni t\mapsto h_t$ of multiplicative reals
$$h_t(x^\zm,y^a)=(x^\zm,t^{w_a}y^a)\,.$$
In this sense, vector bundles are graded bundles of degree $1$, with the Euler vector field as the weight vector field.


\begin{example} As canonical examples of graded bundles can serve
the higher tangent bundles $\sT^kM$ with the adapted coordinates $(x,\dot x, \ddot x,\dots)$ of degrees $0,1,2,\dots$, respectively.
\end{example}
\begin{example}
If $\zt:E\to M$, with affine coordinates $(x^\zm,y^a)$, is a vector bundle, then $\we^2\sT E$ is canonically a graded bundle of degree $2$ with respect to the projection
$$\we^2\sT\zt:\we^2\sT E\to \we^2\sT M\,.$$
\vskip-.5cm\item The adapted coordinates $(x^\zr, y^a,{\dot x}^{\zm\zn}, y^{\zs b}, z^{cd} )$, ${\dot x}^{\zm\zn}=-{\dot x}^{\zn\zm}$, $z^{cd}=-z^{dc}$, coming from the decomposition of a bivector
$$\wedge ^2\sT \sT M\ni u = \frac{1}{2} {\dot x}^{\zm\zn} \frac{\partial}{\partial x^\zm}\wedge \frac{\partial }{\partial x^\zn} + y^{\zs b}\frac{\partial }{\partial x^\zs}\wedge \frac{\partial }{\partial y^b} +\frac{1}{2} {z}^{cd} \frac{\partial }{\partial y^c}\wedge \frac{\partial }{y^d}\,,
$$
are of degrees $0,1,0,1,2$, respectively.


\end{example}


\begin{definition}
Like for \emph{double vector bundles}, two structures of a graded bundle are called \emph{compatible}, i.e. define a \emph{double graded bundle},  if the corresponding weight vector fields commute.
\end{definition}

\begin{example} Our canonical example will be $\we^2\sT E$ with the diagram of bundles and their morphisms
$$
        {\xymatrix@R-5mm @C-3mm{ & \wedge^2\sT E \ar[ld] \ar[rd] & \cr
        \quad E \ar[rd] & & \wedge^2\sT M \ar[ld]  \cr & M  & }}\,.
$$
In particular, $\we^2\sT \we^2\sT^\* M$ is a double graded-vector bundle
$$
        {\xymatrix@R-5mm @C-10mm{ & \wedge^2\sT \we^2\sT^\* M \ar[ld] \ar[rd] & \cr
        \quad \we^2\sT^\* M \ar[rd] & & \wedge^2\sT M \ar[ld]  \cr & M  & }}\,.
$$
\end{example}
On $\wedge ^2 \sT^\*M$, we have the canonical \emph{Liouville 2-form}:
    $$\ezvy^2_M = \frac{1}{2} p_{\zm\zn} \xd x^\zm\wedge \xd x^\zn\,,$$
inducing the  canonical \emph{multisymplectic form}
    $$\ezw^2_M = \xd \ezvy_M^2 =   \frac{1}{2} \xd p_{\zm\zn} \wedge\xd x^\zm\wedge \xd x^\zn\,.$$
The multisymplectic form, \emph{via} the contraction, induces in turn the double graded bundle morphism
    \beas
        \ezb^2_M &\colon  \wedge ^2\sT \wedge^2 \sT^\* M \rightarrow \sT^\*
        \wedge ^2 \sT^\* M  \\
        &\colon u \mapsto \xi_u \ezw^2_M\,.
    \eeas
In local coordinates (summation convention used),
    $$
        \ezb^2_M(x^\zm, p_{\zl\zk},{\dot x}^{\zn\zs},y_{\theta\zr}^\eta,\dot p_{\gamma\delta\epsilon\zeta}) =
        (x^\zm, p_{\zl\zk}, -y_{\eta\zr}^\eta, {\dot x}^{\zn\zs})\,.
    $$
Using now the canonical isomorphism of double vector bundles
$$\cR=\cR_{\wedge^2 \sT M}:\sT^\* \wedge^2 \sT^\* M\to\sT^\* \wedge^2 \sT M\,,$$
we can define $\za_M^2=\cR\circ\zb_M^2$,
 which is another double graded bundle morphism,
    $$ \eza^2_M \colon \wedge ^2\sT \wedge ^2 \sT^\* M  \rightarrow   \sT^\* \wedge ^2\sT M\,,$$
of double graded bundles over $\wedge ^2\sT M$ and $\wedge ^2\sT^\* M$.
In local coordinates,
$$
        \eza^2_M(x^\zm, p_{\zl\zk},{\dot x}^{\zn\zs},y_{\theta\zr}^\eta,\dot p_{\gamma\delta\epsilon\zeta}) =
        (x^\zm, {\dot x}^{\zn\zs}, y_{\eta\zr}^\eta,p_{\zl\zk})\,.
$$
The map $\za_M^2$ can also be obtained as a certain `dual' of the canonical isomorphism
$$\zk_M^2:\sT\we^2\sT M\to\we^2\sT\sT M\,,$$
generalizing the canonical flip (\ref{kappa}).
 Combining the maps $\ezb^2_M$ and $\eza_M^2$, we get the following \emph{Tulczyjew triple} for multivector bundles, consistsing of double graded bundle morphisms:
    $${\xymatrix@R-4mm @C-13mm{ &  \sT^\*\wedge^2\sT^\*M  \ar[ldd]_*{} \ar[rd]^*{} & & & \wedge^2 \sT \wedge^2\sT^\* M \ar[rrr]^*{{\za}^2_M}
    \ar[lll]_*{\zb^2_M} \ar[ldd]^*{} \ar[rd]^*{}& & & \sT^\*\wedge^2\sT M \ar[ldd]^*{} \ar[rd]^*{} & \cr
    & & \wedge^2\sT M \ar[ldd]^*{} & & & \wedge^2\sT M  \ar[ldd]^*{} \ar[lll]^*{} \ar[rrr]^*{} & & & \wedge^2\sT M \ar[ldd]^*{}  \cr
    \wedge^2\sT^\* M  \ar[rd]^*{}  & & & \wedge^2\sT^\* M \ar[rrr]^*{} \ar[lll]^*{} \ar[rd]^*{}& & & \wedge^2\sT^\* M
    \ar[rd]^*{} & & \cr
    & M  & & &  M \ar[rrr]^*{} \ar[lll]^*{} & & & M & }}\,.
    $$

 We have a straightforward generalization for all integer $n\ge 1$ replacing $2$:
$${\xymatrix@R-4mm @C-12mm{ &  \sT^\*\wedge^n\sT^\*M  \ar[ldd]_*{} \ar[rd]^*{} & & & \wedge^n \sT \wedge^n\sT^\* M \ar[rrr]^*{{\za}^n_M}
    \ar[lll]_*{\zb^n_M} \ar[ldd]^*{} \ar[rd]^*{}& & & \sT^\*\wedge^n\sT M \ar[ldd]^*{} \ar[rd]^*{} & \cr
    & & \wedge^n\sT M \ar[ldd]^*{} & & & \wedge^n\sT M  \ar[ldd]^*{} \ar[lll]^*{} \ar[rrr]^*{} & & & \wedge^n\sT M \ar[ldd]^*{}  \cr
    \wedge^n\sT^\* M  \ar[rd]^*{}  & & & \wedge^n\sT^\* M \ar[rrr]^*{} \ar[lll]^*{} \ar[rd]^*{}& & & \wedge^n\sT^\* M
    \ar[rd]^*{} & & \cr
    & M  & & &  M \ar[rrr]^*{} \ar[lll]^*{} & & & M & }}\,.
    $$
The map $\ezb^n_M$,
 \beas
        \ezb^n_M &\colon  \wedge ^n\sT \wedge^n \sT^\* M \rightarrow \sT^\*
        \wedge ^n \sT^\* M\,,  \\
        &\colon u \mapsto \xi_u \ezw^n_M\,,
    \eeas
comes from the canonical multisymplectic $(n+1)$-form $\ezw^n_M=\xd \ezvy_M^n$ on $\wedge ^n \sT^\* M$, being the differential of the canonical Liouville $n$-form $$\ezvy_M^n=\frac{1}{n!}p_{\zm_1\dots,\zm_n}\xd x^{\zm_1}\we\cdots\we\xd x^{\zm_n}\,.$$
The map ${\za^n_M}$ is just the composition of ${\zb^n_M}$ with the canonical isomorphism of double vector bundles $\sT^\*\wedge ^n \sT^\* M$ and $\sT^\*\wedge ^n \sT M$.

\section{String dynamics}
 The way of obtaining the implicit phase dynamics $\cD$ as a submanifold of $\wedge^2 \sT \wedge^2\sT^\* M$, from a Lagrangian $L:\wedge^2\sT M\to\R$ or from a Hamiltonian $H:\wedge^2\sT^\* M\to\R$, is now fully analogous to the case of the standard Tulczyjew triple.
For the Lagrangian, the corresponding diagram reads
{$$\hskip-1cm\xymatrix@C-30pt@R-12pt{
{\mathcal{D}}\ar@{ (->}[r]& \we^2\sT\we^2\sT^\ast M \ar[rrr]^{\alpha_M^2} \ar[dr] \ar[ddl]
 & & & \sT^\ast\we^2\sT M\ar[dr]\ar[ddl] & \\
 & & \we^2\sT M\ar@{.}[rrr]\ar@{.}[ddl]
 & & & \we^2\sT M \ar@{.}[ddl]\ar@/_1pc/[ul]_{\xd L}\ar[dll]_{\cP L}\\
 \we^2\sT^\ast M\ar@{.}[rrr]\ar@{.}[dr]
 & & & \we^2\sT^\ast M\ar@{.}[dr] & &  \\
 & M\ar@{.}[rrr]& & & M &
}\qquad$$}
and
{$$\mathcal{D}=(\alpha_M^2)^{-1}(\xd L(\we^2\sT M))$$}
In local coordinates,
{$$\mathcal{D}=\left\{(x^\zm, p_{\zl\zk},{\dot x}^{\zn\zs},y_{\theta\zr}^\eta,\dot p_{\gamma\delta\epsilon\zeta}):\;\; y_{\eta\zr}^\eta=\frac{\partial L}{\partial x^\zr},\quad p_{\zl\zk}=\frac{\partial L}{\partial \dot{x}^{\zl\zk}}\right\}\,.$$}
On the Hamiltonian side, we get
{$$\hskip-1.2cm\xymatrix@C-25pt@R-12pt{
 & \sT^\ast\we^2\sT^\ast M  \ar[dr] \ar[ddl]
 & & & \we^2\sT\we^2\sT^\ast M\ar[dr]\ar[ddl] \ar[lll]_{\beta_M^2}&
 { \mathcal{D}}\ar@{ (->}[l] \\
 & & \we^2\sT M\ar@{.}[rrr]\ar@{.}[ddl]
 & & & \we^2\sT M \ar@{.}[ddl]\\
 \we^2\sT^\ast M\ar@{.}[rrr]\ar@{.}[dr] \ar@/^1pc/[uur]^{\xd H}
 & & & \we^2\sT^\ast M\ar@{.}[dr] & &  \\
 & M\ar@{.}[rrr]& & & M &
}\qquad$$}
and
{$$\mathcal{D}=(\beta_M^2)^{-1}(\xd H(\we^2\sT^\ast M))\,.$$}
In local coordinates,
{$$\mathcal{D}=\left\{(x^\zm, p_{\zl\zk},{\dot x}^{\zn\zs},y_{\theta\zr}^\eta,\dot p_{\gamma\delta\epsilon\zeta}):\;\; y_{\eta\zr}^\eta=-\frac{\partial H}{\partial x^\zr},\quad {\dot x}^{\zn\zs}=\frac{\partial H}{\partial p_{\zn\zs}}\right\}\,,$$}
that explicitly produces the Hamilton equations.
Again, this framework can be extended to Morse families.

\section{The nonholonomic Euler-Lagrange equations for strings}
To define \emph{nonholonomic Euler-Lagrange equations} for strings, consider
a surface
$$S:\R^2\ni(t,s)\mapsto (x^\zs(t,s))$$
in $M$ and its bi-tangent prolongation
$$\we^2\st S:\R^2\to\we^2\sT M\,,\quad\we^2\st S=\st_tS\we\st_sS\,,
$$
where $\st_t$ and $\st_s$ are tangent prolongations of curves with respect to variables $t$ and $s$, respectively.

It is easy to see that both parameterized surfaces, $\xd L\circ\we^2\st S$ and $\za^2_M\circ\we^2\st(\cP L\circ\we^2\st S)$ in $\sT^*\we^2\sT M$ cover $\we^2\st S$. Therefore, their difference makes sense and, as easily seen, takes values in the annihilator $V^0\we^2\sT M$ of the vertical subbundle $V\we^2\sT M\subset\sT\we^2\sT M$. Since $V^0\we^2\sT M\simeq\we^2\sT M\ti_M\sT^*M$, we obtain a map
$\zd L_S:\R^2\to\sT^*M$. The above map is interpreted as external forces along the string trajectory $S$. Its value at $(t,s)$ depends on the second jet $\sj^2 S(t,s)$ of $S$ only, so defines the variation of the Lagrangian understood as a map
\be\label{works}\zd L:\sJ^2_0(\R^2,M)\to\sT^*M\,,
\ee
where $\sJ^2_0(\R^2,M)$ is the bundle of all second jets of maps $\R^2\to M$ at $0\in \R^2$.
The equation
\be\label{ELs}\zd L_S=0
\ee
we will call the \emph{Euler-Lagrange equation}. It tells that the surface $\xd L\circ\we^2\st S$ corresponds \emph{via} $\za^2_M$ to an \emph{admissible surface} in $\we^2\sT\we^2\sT^*M$, i.e. the bi-tangent prolongation of a parameterized surface in $\we^2\sT^*M$.

If now, $A\subset\we^2\sT M$ is an affine subbundle of $\we^2\sT M$, $\sv(A)$ is the linear part of $A$, and $\sv(A)^0\subset\sT^*M$ its annihilator defined as the set
\be\label{annih}
\sv(A)^0=\{\zh_x\in\sT^*M\, |\, i_{\zh_x}u_x=0\ \text{for all}\ u_x\in\sv(A)_x\}\,,
\ee
then we can replace (\ref{EL}) with
\be\label{dAs}\zd L_S\in \sv(A)^0\,,
\ee
which is the \emph{d'Alembert principle}. It tells that the forces $\zd L_S$ belong to $\sv(A)^0$, so make no work along the trajectory. Indeed, that a one form $\zh$ vanishes on the surface $S$ is  equivalent to the fact that its contraction with $X\we Y$ is zero, for any vector fields $X,Y$, tangent to $S$.

In other words, $S$ satisfies the Euler-Lagrange equations  if the image by $\xd L$  of its prolongation to $\we^2\sT M$,
$$(t,s)\mapsto \left(x^\zs(t,s)),\dot x^{\zm\zn}(t,s)=\frac{\pa x^\zm}{\pa t}\frac{\pa x^\zn}{\pa s}-\frac{\pa x^\zm}{\pa s}\frac{\pa x^\zn}{\pa t}\right)\,,$$
is $\za_M^2$-related to the prolongation of the surface $\mathcal{P}L\circ \we^2\sT S$, living in the phase space $\sT^\ast M$, to $\we^2\sT\we^2\sT^\ast M$.

The constrained Euler-Lagrange equation associated with the affine nonholonomic constraint represented by $A$ take the form
\bea\label{cELs}
&&\we^2\st S\in\sv(A)\,;\\
&&\zd L_S\in \sv(A)^0\,.\label{cELs1}
\eea
In coordinates, they read
\bea\label{nhELs}
&& \left(\frac{\pa S^\zm}{\pa t}\frac{\pa S^\zn}{\pa s}-\frac{\pa S^\zn}{\pa t}\frac{\pa S^\zm}{\pa s}-a^{\zm\zn}\right)\left(\zh^i_\zm\circ S\right)=0\,;\\
&&\frac{\pa L}{\pa x^\zn}-\frac{\pa S^\zm}{\pa t}\frac{\pa}{\pa s}\left(\frac{\pa L}{\pa \dot x^{\zm\zn}}\right)+\frac{\pa S^\zm}{\pa s}\frac{\pa}{\pa t}\left(\frac{\pa L}{\pa \dot x^{\zm\zn}}\right)=\zl_i\zh^i_\zn\,,\label{nhELs1}
\eea
where, as before, $\zh^i=\zh^i_\zn(x)\xd x^\zn$, $i=1,\dots,\dim(\sv(A)^0)$, are one forms generating $\sv(A)^0$, $a=(x^\zs,a^{\zm\zn}(x))$ is an arbitrary section of the affine bundle $A\to M$, and $\zl_i(x)$ are arbitrary coefficient functions.


\begin{example}
 In the dynamics of strings, the manifold of infinitesimal
configurations  is $\wedge ^2 \sT M$, where $M$ is the space time with the
Lorentz metric $g$.  This metric induces a scalar product $h$ in fibers of
$\wedge ^2 \sT M$, so that for
    $$ w=\frac{1}{2} {\dot x}^{\zm\zn}\frac{\partial }{\partial
    x^\zm}\wedge \frac{\partial }{\partial x^\zn}\,,\ \
    u=\frac{1}{2} {\dot x'}{}^{\zm\zn}\frac{\partial }{\partial
    x^\zm}\wedge \frac{\partial }{\partial x^\zn}\,,
    $$
we have
    $$
        (u|w) = h_{\zm\zn\zk\zl} {\dot x}^{\zm\zn}{\dot x'}{}^{\zk\zl}\,,$$
         where

\vskip-.7cm
    $$h_{\zm\zn\zk\zl} = g_{\zm\zk} g_{\zn\zl} - g_{\zm\zl}g_{\zn\zk} \,.$$
 The Lagrangian is a function of the volume with respect to this metric, the so called \emph{Nambu-Goto Lagrangian},
    $$
        L(w) =\sqrt{(w|w)} = \sqrt{h_{\zm\zn\zk\zl} {\dot x}^{\zm\zn}{\dot x}{}^{\zk\zl}}\,,
    $$
which is defined on the open submanifold of positive bivectors.

 The dynamics
    $ {\cD} \subset \wedge ^2\sT \wedge^2 \sT^\* M$
is the inverse image by $\eza^2_M$ of the lagrangian submanifold $\xd L(\we^2\sT M)$ and it is
described by the Lagrange (phase) equations
     \beas
        y^\za_{\za\zn}&= \frac{1}{2\zr} \frac{\partial
        h_{\zm\zk\zl\zs}}{\partial x^\zn} {\dot x}^{\zm\zk} {\dot x}^{\zl\zs},   \\
                       p_{\zm\zn} &=\frac1\zr h_{\zm\zn\zl\zk}{\dot x}^{\zl\zk}\,,                      \eeas
 where
$$\zr = \sqrt{h_{\zm\zn\zl\zk}{\dot x}^{\zm\zn}{\dot x}^{\zl\zk}}\,.$$
 The dynamics $\cD$ is also the inverse image by $\ezb^2_M$ of the lagrangian submanifold in $\sT^\*\wedge^2\sT^\* M$, generated  by the Morse family
    \beas
        H&\colon \wedge^2\sT^\* M\times \R_+ \rightarrow \R\,, \\
        &\colon (p,r)\mapsto r(\sqrt{(p|p)} -1)\,.
    \eeas
\end{example}
\begin{example}
 In the case of a minimal surface, i.e. \emph{the Plateau problem}, we replace the Lorentz metric with a positively defined one.

In particular, if $M=\R^3=\{(x^1=x,x^2=y,x^3=z)\}$ with the Euclidean metric, the Lagrangian reads
$$L(x^\zm,\dot x^{\zk\zl})=\sqrt{\sum_{\zk,\zl}\left(\dot x^{\zk\zl}\right)^2}\,.
$$
 The Euler-Lagrange equation for surfaces
 \be\label{S}S:(x,y)\mapsto (x,y,z(x,y))
  \ee
 provides the well-known equation for minimal surfaces, found already by Lagrange,

\be\label{Lag}\frac{\pa}{\pa x}\left(\frac{z_x}{\sqrt{1+z_x^2+z_y^2}}\right)+\frac{\pa}{\pa y}\left(\frac{z_y}{\sqrt{1+z_x^2+z_y^2}}\right)=0\,.
\ee
 In another form,
$$(1+z_x^2)z_{yy}-2z_xz_yz_{xy}+(1+z_y^2)z_{xx}=0\,.
$$
\end{example}
\begin{example}
For the Plateau problem as above, let us put nonholonomic constraints given by the affine bundle
$$A=\{ \pa_{x}\we\pa_y+f(x,y,z)(\pa_x-\pa_y)\we\pa_z\,|\, f\in C^\infty(\R^3)\}\,.
$$
It can be easily seen that $\sv(A)^0$ is generated by the one-form $\zh=\xd x+\xd y$.
It is the matter of direct computations to show that the surface (\ref{S}) satisfies (\ref{nhELs})
and (\ref{nhELs1}) if and only if (\ref{Lag}) is satisfied together with $z_x=z_y$. But the latter means that $z(x,y)=F(x+y)$ for some function $F$ of one variable, and we end up with the equation
$$\left(\frac{F'}{\sqrt{1+2(F')^2}}\right)'=0\,,
$$
which is equivalent to $F''=0$. Hence, the solutions of our nonholonomic string equation are just planes being the graphs of linear functions $u(x,y)=a(x+y)+b$.
\end{example}


\end{document}